\documentclass{iopart}
\usepackage[T1]{fontenc}

 \expandafter\let\csname equation*\endcsname\relax
 \expandafter\let\csname endequation*\endcsname\relax

\usepackage{amsmath}
\usepackage{amssymb}
\usepackage{graphicx}
\usepackage{esint}



\def \beq {\begin{equation}}
\def \edq {\end{equation}}
\def \bes {\begin{subequations}}
\def \eds {\end{subequations}}
\def \beqn {\begin{equation*}}
\def \edqn {\end{equation*}}

\def \dag {\dagger}

\def \up {\uparrow}
\def \down {\downarrow}

\def \calh {{\cal{H}}}

\def \ket {\rangle}
\def \bra {\langle}

\begin{document}

\title{A hybrid superconducting quantum dot acting as an efficient charge and spin Seebeck diode}

\author{Sun-Yong Hwang$^{1}$, David S\'anchez$^{1}$ and Rosa L\'opez$^{1}$}
\address{$^1$Institut de F\'isica Interdisciplin\`aria i Sistemes Complexos IFISC (UIB-CSIC), E-07122 Palma de Mallorca, Spain}
\ead{david.sanchez@uib.es}

\begin{abstract}
We propose a highly efficient thermoelectric diode device built from the coupling of a quantum dot
with a normal or ferromagnetic electrode and a superconducting reservoir.
The current shows a strongly nonlinear behavior in the forward direction (positive thermal gradients)
while it almost vanishes in the backward direction (negative thermal gradients). Our discussion is supported
by a gauge-invariant current-conserving transport theory accounting for electron-electron interactions
inside the dot. We find that the diode behavior is greatly tuned with external gate potentials, Zeeman splittings
or lead magnetizations. Our results are thus relevant for the search of novel thermoelectric devices with enhanced functionalities.
\end{abstract}

\maketitle

\section{Introduction}
Diodes are building blocks in modern electronics industry due to its ability to show
unidirectional current flow. Thus, in semiconductor p-n junctions the current  $I$ becomes
a non-odd function of the applied voltage $V$, $I(V)\neq -I(-V)$, leading to substantial rectification.
Recently, the interest has shifted to finding diode effects in devices in the presence of a thermal gradient $\theta$~\cite{li04},
$I(\theta)\neq -I(-\theta)$. This is a thermoelectric phenomenon and thereby the name of Seebeck diodes.
Furthermore, the spin current can be also rectified as predicted
in the spin Seebeck diodes~\cite{ren13,bor14,bor14b,ren14,fu15}.
Here, the spin current is generated via the experimentally demonstrated
spin Seebeck effect~\cite{uch08,sla10,bau10}.

In quantum coherent conductors coupled to normal metallic leads, the thermoelectric current
becomes strongly nonlinear when the local density of states is energy dependent and more than
one resonance is involved in the transmission function~\cite{fah13,ser14}.
Otherwise, the weakly nonlinear terms in a current--temperature expansion are small
compared to the linear response coefficients~\cite{san13,hwa14}.
These nonlinearities precisely describe, to leading order, rectification and diode effects~\cite{jia15}.
We have recently shown that a quantum dot
sandwiched between ferromagnetic and superconducting terminals 
exhibits large thermoelectric power and figure of merit~\cite{hwa15}.
The effect arises because a spin-split dot level allows for tunneling from the hot
metallic lead to the available quasiparticle states in the cold superconducting side~\cite{oza14,kol15,kal14,mac14}.
Nevertheless, our analysis was valid in the linear regime
of transport only. In this paper, we consider the nonlinear case. Surprisingly, we find a highly efficient
diode effect that works equally well for both the charge and the spin transport flow.
The basic operating principle of our device relies on a strong energy dependence of the transmission function which naturally arises in the quasiparticle spectrum of normal-superconducting junction.

A careful calculation of the current--voltage characteristics beyond linear response
requires knowledge of the nonequilibrium screening potential inside the mesoscopic structure~\cite{but93}.
When the nanosystem is subjected to the application of large thermal gradients, one needs to determine
the variation of the internal electrostatic field to temperature shifts~\cite{san13,whi13,mea13}.
For large quantum dots or for dots strongly coupled to the leads (weak Coulomb blockade regime~\cite{bro05}),
it suffices to treat electron-electron interactions at the mean-field level. We consider a single-level dot 
with fluctuating potential $U$ due to injected charges from the attached leads, see Fig.~1.
A recent work reports the observation of weak diode effects in a superconductor coupled to a two-dimensional
electron gas~\cite{lo13}. We here propose that a hybrid quantum dot working as an energy filter between
the normal reservoir and the superconducting terminal~\cite{gra04,dea10} leads to much stronger diode features
with rectification efficiencies close to unity.

\begin{figure}[t]
\centering
  \begin{centering}
    \includegraphics[width=0.55\textwidth,clip]{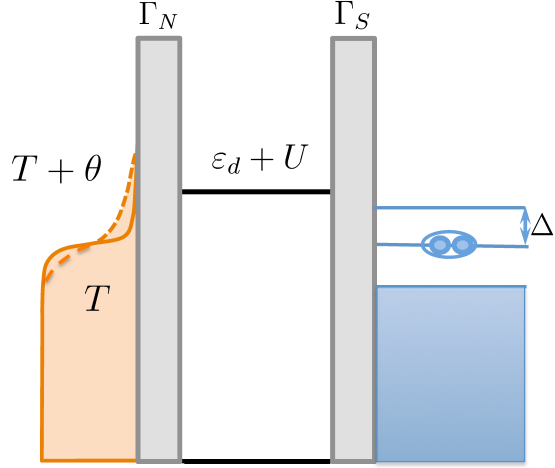}
  \end{centering}
  \caption{Sketch of our Seebeck diode. Left normal (N) or ferromagnetic (F) lead can be heated or cooled, which respectively generates thermal broadening (dashed orange line) or sharpening (full orange line) of the Fermi function. The right superconductor (S) maintains the thermal equilibrium. As a consequence, at low background temperature $T$ the states below the gap are filled (blue color). The energy level $\varepsilon_d$ of the quantum dot sandwiched between tunnel barriers (gray color) of transparencies $\Gamma_N$ and $\Gamma_S$ can be renormalized by interaction $U$ and tunable by a back gate potential away from the Fermi energy (blue line). The potential $U$ shifts upward as the forward thermal bias ($\theta>0$) is applied creating a synergetic effect on the strongly nonlinear current with the thermally excited quaisiparticles from the left lead. On the other hand, cooling with a backward thermal bias ($\theta<0$) lowers the current as the number of available states sharply decreases.}
  \label{fig:sketch}
\end{figure}

\section{Formalism}
Our Seebeck diode consists of a ferromagnetic (F) reservoir characterized by a spin-polarization $p$ ($|p|\le1$), a single-level quantum dot (D), and the superconductor (S), as depicted in Fig.~1. The normal metal case (N) has equal spin up and down densities, we therefore put $p=0$ in the left lead. We write the model Hamiltonian~\cite{cao04}
\beq\label{H}
\calh=\calh_{L}+\calh_{S}+\calh_{D}+\calh_{T}\,,
\edq
where
\beq
\calh_{L=N,F}=\sum_{k\sigma}\varepsilon_{Lk\sigma}c_{Lk\sigma}^{\dag}c_{Lk\sigma}
\edq
describes the left N or F lead with charge carriers of momentum $k$, spin $\sigma=\up,\down$, and energy $\varepsilon_{Lk\sigma}$, and
\beq
\calh_{S}=\sum_{k\sigma}\varepsilon_{Sk\sigma}c_{Sk\sigma}^{\dag}c_{Sk\sigma}+\sum_{k}\big[\Delta c_{Sk\up}^{\dag}c_{S,-k\down}^{\dag}+{\rm H.c.}\big]
\edq
is the superconductor Hamiltonian with the energy gap $\Delta$. We consider an equilibrium superconductor where the phase of $\Delta$ can be neglected by a gauge transformation, hence void of AC Josephson effect arising from the phase evolution. Importantly, in the dot Hamiltonian of Eq.~\eqref{H}
\beq\label{HD}
\calh_{D}=\sum_{\sigma}(\varepsilon_{d\sigma}+U_\sigma)d_{\sigma}^{\dag}d_{\sigma}\,,
\edq
the spin-dependent energy level $\varepsilon_{d\sigma}=\varepsilon_d+\sigma\Delta_Z$ is renormalized by the internal potential $U_\sigma$ that accounts for the Coulomb interaction. The Zeeman splitting $\Delta_Z$ is finite when the magnetic field is on.
The screening potential $U=\sum_\sigma U_\sigma$ is determined by solving the Poisson's equation which for homogeneous potentials reads $\delta q=q-q_\text{eq}=C(U-V_g)$ where $C$ and $V_g$ are the capacitance of the dot and the gate potential applied to it, respectively. We consider the charge neutral limit ($C=0$), an experimentally relevant situation for strongly interacting dots. 
The solution can be expressed by the lesser Green's function~\cite{wan01}, i.e., $q=-i\int d\varepsilon~G^<(\varepsilon)$, where $G^<(t,t')=i\bra d^\dag(t')d(t)\ket$.
We also consider the spin-generalized case~\cite{cao04} and solve the Poisson's equation in a spin-dependent manner~\cite{lop12} incorporating the ferromagnet polarization and the magnetic field applied to the quantum dot.
In this case, the spin-dependent charge density reads $q_\sigma=-i\int d\varepsilon~G_\sigma^<(\varepsilon)$ where $G_\sigma^<(\varepsilon)$ is explicitly written in~\ref{appendix}.
In order to take into account full nonlinearity of the temperature gradient $\theta$ [Fig.~1], we numerically solve two nonlinear equations 
\beq\label{deltaq}
\delta q_\up(\theta,U_\up,U_\down)=\delta q_\down(\theta,U_\up,U_\down)=0.
\edq
Here, the screening potential fluctuates in order to keep the dot charge constant.
This gives the solution of the form $U_\sigma=U_\sigma(\theta)$ for each spin $\sigma=\up,\down$ valid to all orders in a temperature expansion of the potential. We find that interactions favor the diode effect. This will be discussed more in detail below.
Finally, the tunneling Hamiltonian in Eq.~\eqref{H} reads
\beq
\calh_{T}=\sum_{k\sigma}t_{L\sigma}c_{Lk\sigma}^{\dag}d_{\sigma}+\sum_{k\sigma}t_{S\sigma}c_{Sk\sigma}^{\dag}d_{\sigma}+{\rm H.c.}\,,
\edq
where $t_{\alpha\sigma}$ is the hopping amplitude between the quantum dot and each lead $\alpha=L,S$. 

The spin-resolved current $I_\sigma=-(ie/\hbar)\bra[\calh,N_{L\sigma}]\ket$ can be evaluated from the time evolution of electron number $N_{L\sigma}=\sum_{k}c_{Lk\sigma}^{\dag}c_{Lk\sigma}$ in the left lead by employing the nonequilibrium Keldysh Green's function technique~\cite{Cue96,Sun99}. In the isoelectric case with no voltage bias $V=0$, the subgap Andreev current is completely blocked since $I_A^\sigma=(e/h)\int d\varepsilon T_A^\sigma(\varepsilon)[f_L(\varepsilon-eV)-f_L(\varepsilon+eV)]$ is identically zero to all orders in $\theta$~\cite{hwa15b}.
This insensitivity of $I_A^\sigma$ to thermal gradients only is a manifestation of the particle-hole symmetry inherent in the subgap transport.
Consequently, the total current emerges only from the quasiparticle contribution; hence we can write the spin-resolved current
\begin{equation}
I_{\sigma}=\frac{e}{h}\int d\varepsilon~T_{Q}^{\sigma}(\varepsilon)\big[f_{L}(\varepsilon)-f_{S}(\varepsilon)\big]\,,\label{I_sig}
\end{equation}
where $f_{\alpha=L,S}(\varepsilon)=\{1+\exp[(\varepsilon-E_F)/k_{B}T_{\alpha}]\}^{-1}$ is the Fermi-Dirac distribution function with local temperature for each lead $T_\alpha=T+\theta_\alpha$ ($T$: background temperature, $\theta_\alpha$: thermal bias). We apply the thermal gradient $\theta$ only to the left non-superconducting lead ($T_L=T+\theta$) while the superconductor maintains the equilibrium temperature $T_S=T$ ($\theta_S=0$) and take the Fermi level to be $E_F=0$. Thus, the forward thermal bias is defined by $\theta>0$ and the backward one by $-T<\theta<0$.

Importantly, the quasiparticle transmission in Eq.~\eqref{I_sig} is proportional to the superconducting density of states $\Theta(|\varepsilon|-\Delta)/\sqrt{\varepsilon^{2}-\Delta^{2}}$, i.e.,
\beq\label{TQ}
T_{Q}^{\sigma}(\varepsilon)\propto\Gamma_{L\sigma}\Gamma_{S}\frac{\Theta(|\varepsilon|-\Delta)|\varepsilon|}{\sqrt{\varepsilon^{2}-\Delta^{2}}}\,,
\edq
where $\Gamma_{L\sigma}=\Gamma_L(1+\sigma p)=2\pi|t_{L\sigma}|^2 \sum_k \delta (\varepsilon-\varepsilon_{Lk\sigma})$ and $\Gamma_S=2\pi|t_{S\sigma}|^2\sum_{k}\delta(\varepsilon-\varepsilon_{Sk\sigma})$ are the tunnel broadenings to each lead in the wide-band approximation, and $\Theta(\varepsilon)$ is the Heaviside step function, respectively.
An explicit expression of $T_{Q}^{\sigma}(\varepsilon)$ can be found in~\ref{appendix}.
It clearly follows from Eq.~\eqref{TQ} that due to the energy gap $\Delta$ of the superconducting lead, one needs to apply high enough (forward) thermal bias to the system in order to activate the quasiparticle contribution. On the other hand, the quasiparticle current can be deactivated when we cool the system down, i.e., applying backward thermal gradient with $\theta<0$, in which case the current is highly suppressed. This comprises the working principle of our charge and spin Seebeck diode proposed here: (i) complete suppression of the parasitic Andreev current with $V=0$, (ii) activation of quasiparticles above the superconducting gap with the forward temperature gradient $\theta>0$ but not the other way round with $\theta<0$.

Now, the combination of superconductivity and spintronics can lead to novel functionalities with better performances~\cite{lin15,lin16}.
In order to realize the spin Seebeck diode~\cite{ren13,bor14,bor14b,ren14,fu15}. either finite magnetic field $\Delta_Z\ne0$ or a nonzero polarization $p\ne0$ using the ferromagnet is necessary to break the spin symmetry of the transmission, viz. $T_Q^\up(\varepsilon)\ne T_Q^\down(\varepsilon)$ in Eq.~\eqref{TQ}. However, even in nonmagnetic case with $p=\Delta_Z=0$, the charge current--temperature curves would clearly show the charge Seebeck diode features owing to the underlying mechanism explained above. 
Below, we discuss $I_c-\theta$ and $I_s-\theta$ characteristics in the isoelectric case where the charge ($I_c$) and spin ($I_s$) currents are defined with the aid of Eq.~\eqref{I_sig}:
\begin{align}
&I_c=I_\up+I_\down\,,\\
&I_s=I_\up-I_\down\,.
\end{align}

\section{Results and discussion}
\begin{figure}[t]
\begin{center}
\includegraphics[width=0.4\textwidth,clip,angle=-90]{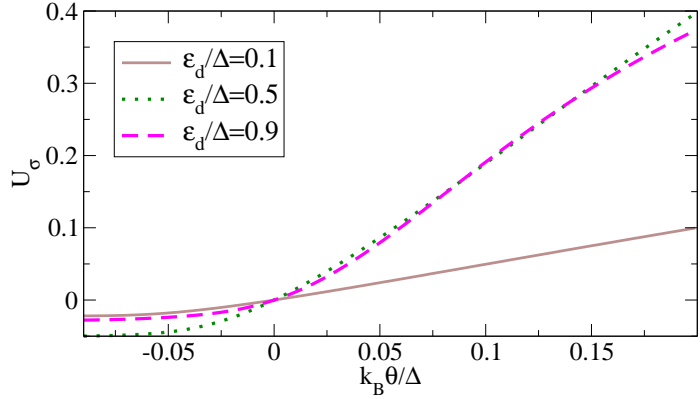}
\caption{$U_\sigma$ versus $\theta$ for several $\varepsilon_d$ at $k_BT=0.1\Delta$ and $p=\Delta_Z=0$ with $\Gamma_{N}\ll\Gamma_S$.}
\label{Fig2}
\end{center} 
\end{figure}

We firstly discuss the interaction effects characterized by the screening potential. Figure~\ref{Fig2} shows $U_\sigma$ as a function of $\theta$ in a N-D-S device where $p=\Delta_Z=0$. The potential $U_\sigma$ for $\theta<0$ is rather suppressed whereas it linearly increases for $\theta>0$. In addition, its linear slope saturates as we increase the dot level beyond $\varepsilon_d=0.5\Delta$ close to the superconductor gap for $\theta>0$ while the potential decreases further for $\theta<0$ as $\varepsilon_d$ approaches $\Delta$. We emphasize that interaction effects are beneficial for the diode behavior discussed here since the forward thermal bias $\theta>0$ shifts the effective dot level higher than that of noninteracting limit to keep the dot charge constant. This is a nice property that clearly makes the synergy with the thermally excited quasiparticle states in the left normal contact.

\begin{figure}[t]
\begin{center}
\includegraphics[width=0.5\textwidth,clip, angle=-90]{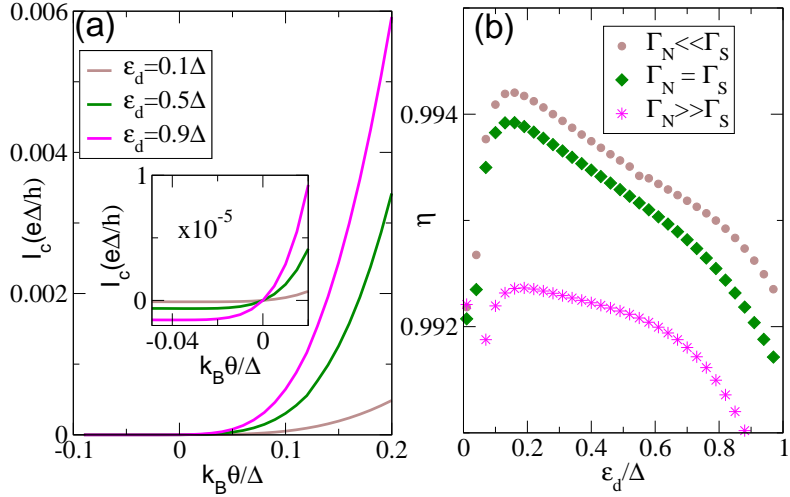}
\caption{(a) $I_c$ versus $\theta$ for several $\varepsilon_d$ at $p=\Delta_Z=0$. The case for $\Gamma_N\ll\Gamma_S$ is shown. (b) $\eta$ versus $\varepsilon_d$ at $k_B\theta_0=0.07\Delta$ for different coupling limits, where $\Gamma_N=0.1\Delta$, $\Gamma_N=0.3\Delta$, and $\Gamma_N=0.5\Delta$ for each case while the total broadening is fixed, i.e., $\Gamma_N+\Gamma_S=0.6\Delta$. The background temperature is $k_BT=0.1\Delta$. Inset of (a) shows that the Ohmic region with $I_c(\theta)=-I_c(-\theta)$ is very narrow.}
\label{Fig3}
\end{center} 
\end{figure}

Figure~\ref{Fig3} displays the charge Seebeck diode behavior of our hybrid device and its high rectification efficiency. For the moment, a purely nonmagnetic case $p=\Delta_Z=0$ in a N-D-S setup is considered.
In Fig.~\ref{Fig3}(a), the charge current for backward thermal gradients $\theta<0$ is greatly suppressed as discussed above whereas strongly nonlinear thermocurrent is generated by heating ($\theta>0$) the normal metallic lead. Moreover, the forward current can be amplified by tuning the gate potential as shown with several dot level positions. $I_c$ increases as the dot level position approaches the superconducting gap onset and it is reinforced by interaction effects.

The rectification efficiency can be quantified by
\beq\label{rect}
\eta=\frac{|I_c(\theta_0)|-|I_c(-\theta_0)|}{|I_c(\theta_0)|}
\edq
for fixed forward and backward thermal gradients $\pm\theta_0$. This number is bounded and the maximum efficiency is given by $\eta=1$ if the backward thermocurrent completely vanishes.
In Fig.~\ref{Fig3}(b), $\eta$ is shown as a function of $\varepsilon_d$ at $k_B\theta_0=0.07\Delta$. This thermal bias is about $250$ mK for Al, still lower than the background temperature. Therefore, we do not need large temperature bias to observe the diode effect [inset of Fig.~\ref{Fig3}(a)]. Remarkably, the rectification is very efficient as $\eta$ is close to unity for various coupling limits, i.e., stronger coupling to S or N and an identical tunnel broadening to each lead. This shows the robustness of our device to unintentional variations of the coupling values to the external contacts. 
Albeit not shown, high efficiencies displayed here are rather insensitive to the change of background temperature $T$.
Another useful way of quantifying the efficiency of our device is to introduce the asymmetry ratio defined by
\beq\label{R}
R=\frac{|I_c(\theta_0)|}{|I_c(-\theta_0)|}.
\edq
One can easily find the relation $R=1/(1-\eta)$ from Eq.~\eqref{rect}. Table~\ref{tab} displays a fast growth of $R$ as a function of $\theta_0$, which can be inferred from Fig.~\ref{Fig3}.

\begin{table}
\caption{Asymmetry ratio $R$ for several $\varepsilon_d$ and $\theta_0$.}
\begin{center}
\begin{tabular}{|c|c|c|c|}
\hline
& $k_B\theta_0=0.01\Delta$ & $k_B\theta_0=0.04\Delta$ & $k_B\theta_0=0.07\Delta$ \\
\hline
$\varepsilon_d=0.1\Delta$ & $2.68$ & $31$ & $166$\\
\hline
$\varepsilon_d=0.5\Delta$ & $2.64$ & $30$ & $155$\\
\hline
$\varepsilon_d=0.9\Delta$ & $2.53$ & $27$ & $134$\\
\hline
\end{tabular}\label{tab}
\end{center}
\end{table}

\begin{figure}[t]
\begin{center}
\includegraphics[width=0.5\textwidth,clip,angle=-90]{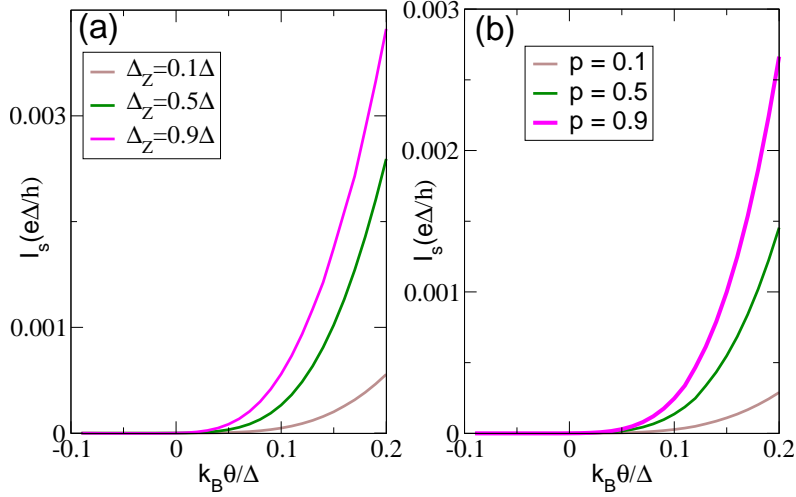}
\caption{$I_s$ versus $\theta$ at (a) $p=0$ for several $\Delta_Z$, and (b) $\Delta_Z=0$ for several $p$. As shown in (a), spin Seebeck diode can be embodied even without a ferromagnetic lead. We have fixed $\varepsilon_d=0.5\Delta$ and $k_BT=0.1\Delta$ with $\Gamma_{N,F}\ll\Gamma_S$.}
\label{Fig4}
\end{center} 
\end{figure}

Figure~\ref{Fig4}(a) shows the spin Seebeck diode feature~\cite{ren13,bor14,bor14b,ren14,fu15} in a N-D-S device with a magnetic field applied to the dot, i.e., $\Delta_Z\ne0$. The ferromagnet is not an essential ingredient if the Zeeman splitting in the dot is nonzero. We observe a quick increase of the spin current as a function of $\theta$. This increase is more dramatic for higher Zeeman splitting because then the dot level allows for greater current into the empty quasiparticle states. In Fig~\ref{Fig4}(b), a F-D-S setup with a nonzero polarization $p\ne0$ also exhibits the spin current rectification depending on the thermal bias direction. In this case, $I_s$ increases for higher $p$ due to more available states with spin up in the source contact. The analogous rectification efficiencies [Eq.~\eqref{rect} but with $I_s(\pm\theta_0)$] for both Figs.~\ref{Fig4}(a) and \ref{Fig4}(b) are also as high as the charge current counterpart (not shown here). Our results suggest that this Seebeck diode device based on the hybrid superconducting quantum dot is very efficient and versatile.

In a realistic superconductor sample, the energy gap depends on the temperature, e.g., $\Delta(T)=\Delta_0\sqrt{1-(T/T_c)^2}$, where $T_c$ is the superconducting critical temperature of the material. If we take Al for a superconductor, its zero temperature energy gap is about $\Delta_0=0.34$ meV with $T_c=1.2$ K. Then, one can easily estimate $\Delta(500~\text{mK})\approx0.9\Delta_0$ with the background temperature $k_BT=0.1\Delta_0$ we have used in this paper. This means that Al superconducting gap is mostly unaffected up to rather high temperatures $T\approx500$ mK. One can therefore practically embody the Seebeck diode as suggested here with, e.g., an Al superconductor and a nanowire or a carbon nanotube quantum dot. A typical current value is $0.001e\Delta/h\approx13$ pA, which is within the reach of today's experimental techniques~\cite{kol15}. For the magnetic configurations, however, $\Delta_Z=0.1\Delta$ corresponds to $B\approx0.03$ T for a nanowire quantum dot with an effective $g$-factor 40. This already exceeds the critical field $B_c=0.01$ T of Al, hence in this case a superconductor with a higher $B_c$, e.g., Nb compounds, should be used to observe the effects shown in Fig.~\ref{Fig4}(a).

\section{Summary}
Since thermoelectric generators and coolers have thus far shown low efficiencies, it is crucial to propose efficient thermoelectric devices with new purposes. Here, we have proposed a proof-of-principle design for a charge and spin Seebeck diode built from the hybrid superconductor quantum dot device. Either normal metallic or ferromagnetic lead can be attached to the quantum dot. Our device shows strong rectification and diode effects as the rectification efficiency is very close to 100$\%$.
We have found that the diode features in the device are highly tunable with back gate potentials, magnetic fields, and lead magnetizations which opens the route for its use in information processing applications.

We have treated Coulomb interactions in the mean-field approximation. In this case, the potential shift is a function of the temperature gradient applied to the non-superconducting lead. Our calculations are valid for metallic dots with good screening properties~\cite{bro05}. We expect that the diode behaviors would survive for a broad range of interaction strengths, even beyond mean field, since the main underlying mechanism of rectification effects is the gapped quasiparticle spectrum with a complete suppression of the subgap transport.

\section*{Acknowledgments}
The authors acknowledge the support from MINECO under Grant No.\ FIS2014-52564 and the Korean NRF under Grant No.\ 2014R1A6A3A03059105.

\appendix
\section{Green's functions and quasiparticle transmission}\label{appendix}
In the isoelectric case with $V=0$, the lesser Green's functions are given by
\beq
\begin{split}
G_{\up}^{<}(\varepsilon)&=\frac{i}{2\pi}f_{L}(\varepsilon)\Big[\Gamma_{L\up}\big|G_{11}^{r}(\varepsilon)\big|^{2}+\Gamma_{L\down}\big|G_{12}^{r}(\varepsilon)\big|^{2}\Big]\\
&\qquad
+\frac{i\widetilde{\Gamma}_{S}}{2\pi}f_{S}(\varepsilon)\Big[\big|G_{11}^{r}(\varepsilon)\big|^{2}
	+\big|G_{12}^{r}(\varepsilon)\big|^{2}-\frac{2\Delta}{|\varepsilon|}
	\text{Re}\big[G_{11}^{r}(\varepsilon)G_{12}^{r,*}(\varepsilon)\big]\Big]\label{G_up^<}\,,
\end{split}
\edq
\beq
\begin{split}
G_{\down}^{<}(\varepsilon)&=\frac{i}{2\pi}f_{L}(\varepsilon)\Big[\Gamma_{L\down}\big|G_{33}^{r}(\varepsilon)\big|^{2}+\Gamma_{L\up}\big|G_{34}^{r}(\varepsilon)\big|^{2}\Big]\\
&\qquad
+\frac{i\widetilde{\Gamma}_{S}}{2\pi}f_{S}(\varepsilon)\Big[\big|G_{33}^{r}(\varepsilon)\big|^{2}
	+\big|G_{34}^{r}(\varepsilon)\big|^{2}+\frac{2\Delta}{|\varepsilon|}
	\text{Re}\big[G_{33}^{r}(\varepsilon)G_{34}^{r,*}(\varepsilon)\big]\Big]\label{G_down^<}\,,
\end{split}
\edq
where $\Gamma_{L\sigma}=\Gamma_L(1+\sigma p)$ and $\widetilde{\Gamma}_{S}=\Gamma_{S}\Theta(|\varepsilon|-\Delta)|\varepsilon|/\sqrt{\varepsilon^{2}-\Delta^{2}}$. Then, the spin-generalized charge fluctuations in Eq.~\eqref{deltaq} can be written as
\begin{align}
&\delta q_\up=-i\int d\varepsilon\big[G_{\up}^{<}(\varepsilon)-G_{\up,\text{eq}}^{<}(\varepsilon)\big]\,,\\
&\delta q_\down=-i\int d\varepsilon\big[G_{\down}^{<}(\varepsilon)-G_{\down,\text{eq}}^{<}(\varepsilon)\big]\,,
\end{align}
for each spin, respectively, where $G_{\sigma,\text{eq}}^{<}(\varepsilon)$ is the value of $G_{\sigma}^{<}(\varepsilon)$ at thermal equilibrium.
The retarded Green's functions which we have used in the above expressions are explicitly given by
\begin{align}
&G^{r}_{11}(\varepsilon)=\Big[\varepsilon-\widetilde{\varepsilon}_{d\up}+\frac{i\Gamma_{L\up}}{2}+\frac{i\Gamma_{S}}{2}\beta_d(\varepsilon)+\frac{\Gamma_S^{2}\Delta^{2}}{4(\varepsilon^2-\Delta^{2})}A_1^r(\varepsilon)\Big]^{-1},\\
&G^{r}_{33}(\varepsilon)=\Big[\varepsilon-\widetilde{\varepsilon}_{d\down}+\frac{i\Gamma_{L\down}}{2}+\frac{i\Gamma_{S}}{2}\beta_d(\varepsilon)+\frac{\Gamma_S^{2}\Delta^{2}}{4(\varepsilon^2-\Delta^{2})}A_2^r(\varepsilon)\Big]^{-1},\\
&G^{r}_{12}(\varepsilon)=G^{r}_{11}(\varepsilon)\frac{i\Gamma_{S}}{2}\beta_o(\varepsilon)A_1^r(\varepsilon)\,,\\
&G^{r}_{34}(\varepsilon)=-G^{r}_{33}(\varepsilon)\frac{i\Gamma_{S}}{2}\beta_o(\varepsilon)A_2^r(\varepsilon)\,,
\end{align}
with
\begin{align}
&A_1^r(\varepsilon)=\Big[\varepsilon+\widetilde{\varepsilon}_{d\down}+\frac{i\Gamma_{L\down}}{2}+\frac{i\Gamma_{S}}{2}\beta_d(\varepsilon)\Big]^{-1},\\
&A_2^r(\varepsilon)=\Big[\varepsilon+\widetilde{\varepsilon}_{d\up}+\frac{i\Gamma_{L\up}}{2}+\frac{i\Gamma_{S}}{2}\beta_d(\varepsilon)\Big]^{-1},\\
&\beta_d(\varepsilon)=\frac{\Theta(|\varepsilon|-\Delta)|\varepsilon|}{\sqrt{\varepsilon^{2}-\Delta^{2}}}-i\frac{\Theta(\Delta-|\varepsilon|)\varepsilon}{\sqrt{\Delta^{2}-\varepsilon^{2}}}\,,\\
&\beta_o(\varepsilon)=\frac{\Theta(|\varepsilon|-\Delta)\text{sgn}(\varepsilon)\Delta}{\sqrt{\varepsilon^{2}-\Delta^{2}}}-i\frac{\Theta(\Delta-|\varepsilon|)\Delta}{\sqrt{\Delta^{2}-\varepsilon^{2}}}\,,
\end{align}
where $\widetilde{\varepsilon}_{d\sigma}=\varepsilon_{d\sigma}+U_\sigma$ represents the renormalized quantum dot level by spin-dependent interaction $U_\sigma$ [see Eq.~\eqref{HD}].

With explicit expressions for the retarded Green's functions, the spin-dependent quasiparticle transmission in Eq.~\eqref{TQ} is given by
\begin{align}
&T_{Q}^{\up}(\varepsilon)=\Gamma_{L\up}\widetilde{\Gamma}_{S}\Big(\big|G_{11}^{r}(\varepsilon)\big|^{2}+\big|G_{12}^{r}(\varepsilon)\big|^{2}-\frac{2\Delta}{|\varepsilon|}\text{Re}\big[G_{11}^r(\varepsilon)G_{12}^{r,*}(\varepsilon)\big]\Big)\,,\\
&T_{Q}^{\down}(\varepsilon)=\Gamma_{L\down}\widetilde{\Gamma}_{S}\Big(\big|G_{33}^{r}(\varepsilon)\big|^{2}+\big|G_{34}^{r}(\varepsilon)\big|^{2}+\frac{2\Delta}{|\varepsilon|}\text{Re}\big[G_{33}^r(\varepsilon)G_{34}^{r,*}(\varepsilon)\big]\Big)\,,
\end{align}
for each spin where $\widetilde{\Gamma}_{S}=\Gamma_{S}\Theta(|\varepsilon|-\Delta)|\varepsilon|/\sqrt{\varepsilon^{2}-\Delta^{2}}$.

\end{document}